\documentclass[twocolumn,epjc3]{svjour3}  
\pdfoutput=1
\smartqed  
\RequirePackage{graphicx}
\RequirePackage{fix-cm}
\RequirePackage{mathptmx}      
\RequirePackage{footnote}
\newcommand \su[1]{\mathrm{SU}(#1)}
\newcommand \tr{\mathrm{Tr}}
\newcommand \LL{\mathrm{L}}
\journalname{Eur. Phys. J. C}
\begin{document}

\title{Topology of Minimal Walking Technicolor
}


\author{Ed Bennett\thanksref{e1,addr1}
        \and
        Biagio Lucini\thanksref{e2,addr1} 
}

\thankstext{e1}{e-mail: pyedward@swan.ac.uk}
\thankstext{e2}{e-mail: B.Lucini@swan.ac.uk}


\institute{Department of Physics, College of Science, Swansea University, Singleton Park, Swansea, SA2 8PP \label{addr1}
}

\date{Received: date / Accepted: date}

\maketitle

\begin{abstract}
We perform a lattice study of the topological susceptibility and instanton size
distribution of the $\su{2}$ gauge theory with two adjoint
Dirac fermions (also known as Minimal Walking Technicolor), which is known
to be in the conformal window. In the theory deformed with a small
mass term, by drawing a comparison with the pure gauge theory, we find
that topological observables are decoupled from the
fermion dynamics. This provides further evidence for the infrared
conformality of the theory. A study of the instanton size distribution
shows that this quantity can be used to detect the onset of finite
size effects.  
\end{abstract}

\section{Introduction}
\label{intro}
After the recent experimental breakthroughs that allowed the announcement of
the discovery of a new bosonic particle with mass at the electroweak
scale~\cite{atlas2012gk,cms2012gu}, the focus of experiments is moving towards the
determination of the nature of this boson, with the goal of finding
hints for new physics. Hence, it is of utmost importance for theorists
to investigate possible scenarios of electroweak symmetry
breaking beyond the standard model. An appealing possibility is the existence of a new strong
interaction whose chiral condensate breaks electroweak
symmetry. Based on this idea, over the years a physically consistent framework
denoted as Walking Techinicolor (WT) was
developed~\cite{Weinberg1975gm,Susskind1978ms,Eichten1979ah,Holdom1981rm,Holdom1984sk,Yamawaki1985zg,Appelquist1986an,Miransky1996pd,Dietrich2006cm,Dietrich2006ck}
(reviews on the topic
include~\cite{Hill2002ap,Piai2010ma,Andersen2011yj}). In
order to be consistent with the stringent electroweak precision measurements, a
theory realising the WT scenario must have two crucial properties:
dynamics characterised by near-conformality, and a chiral condensate
with anomalous dimension of order one. In most 
semi-analytical treatments of WT, both properties are taken as
reasonable assumptions. In order to prove that there exist realistic
gauge theories characterised by near-conformal behaviour and
large anomalous dimension, calculations from first principles are
needed. Gauge-string duality techniques can be used to engineer theories with
the required properties in the limit in which the number of gauge
bosons goes to infinity (see
e.g.~\cite{Nunez2008wi,Elander2009pk,Kutasov2012uq,Levkov2012yk,Anguelova2012ka,Clark2012pm}
and references therein). For a gauge theory based 
on a finite Lie group, lattice computations provide the best
quantitative tool. 

In recent years, a large body of activity has been
devoted by the lattice community to computations aimed at
characterising the phases of gauge theories, with the goal of
understanding signatures of the conformal and near-conformal
regime. The current understanding is summarised
in~\cite{Neil2012cb}. Among the theories investigated to date, $\su{2}$ with 
two Dirac fermions in the adjoint representation (also known as
Minimal Walking
Technicolor~\cite{Sannino:2004qp,Dietrich:2005jn,Foadi:2007ue}, for
which experimental constraints have been recently reported in~\cite{atlas-mwt}) plays an important role:
it is the only theory studied so far to show unambiguously the expected
behaviour of an infrared conformal theory, in independent
numerical results obtained with different techniques and targeting
different observables. Although for building a technicolor model one needs
a near-conformal gauge theory, first principle studies of
infrared conformal theories are interesting both from a
theoretical point of view and from a phenomenological
perspective~\cite{Luty2004ye}. After the first exploratory
calculations of physical quantities in this
theory~\cite{Catterall2007yx,DelDebbio2008zf,Catterall2008qk,Hietanen2008mr},
which were fundamental in ascertaining its conformal behaviour,
lattice studies have focussed on the running of the coupling near the
fixed point~\cite{Hietanen2009az,Bursa2009we,DeGrand2011qd,Catterall2011zf}, on the
calculation of the anomalous dimension of the chiral
condensate~\cite{Lucini2009an,DelDebbio2009fd,DelDebbio2010hu,DelDebbio2010hx,Giedt2012rj,Patella2012da} 
and more recently on controlling finite size
effects~\cite{Karavirta2011mv,Bursa2011ru,DelDebbio2011kp}. The main
goals of this community effort have
been to establish a reliable technique to extract the anomalous
dimension and to identify a good set of observables that could provide
an unambiguous signature of conformal behaviour. These investigations are
expected to be relevant also for the near-conformal case, in which the
theory will mimic the properties of a conformal theory at some
intermediate distances before showing its true confining and chiral
symmetry breaking nature at asymptotically large distances.

To date, the attention of lattice studies has focussed on
spectral quantities. In this paper, we shall report on a first
investigation of topological
quantities\footnote{See~\cite{Nogradi2012dj,deForcrand2012se} for
  studies of topological quantities in a similar context in the $\mathrm{O}(3)$ model.}, and in particular on the
behaviour of the topological charge, of the instanton size distribution
and of the topological susceptibility. These quantities play an
important role in confining and chiral symmetry breaking theories like
QCD (see e.g.~\cite{Delia2012vv} for a recent work), but their
behaviour in (near-)conformal gauge theories has not been explored
before. We shall study whether those observables can be 
used to identify the infinite volume regime and the scaling regime in
the mass-deformed theory. Taking inspiration from the perturbative argument given
in~\cite{Miransky1998dh}, in~\cite{DelDebbio2009fd} it has been shown
from first principles
that $\su{2}$ gauge theory with two Dirac adjoint flavours behaves like a
confining theory with heavy adjoint quarks for any finite value of the
fermion mass, which plays the role of a deforming parameter in a
would-be infrared conformal theory. Hence, in
the infinite volume limit, we expect topological observables to be
compatible with those of Yang-Mills, with any small deviation
originating from the fermionic matter, and large deviations being due to
finite size effects. The decoupling of instantons in the chiral limit
has been discussed also in~\cite{Sannino:2008pz}.

The purpose of this investigation is twofold. On the one hand, we aim
to verify that for large volumes topological variables are broadly
compatible with Yang-Mills values and at characterising qualitatively
their behaviour on smaller volumes. On the other hand, we want to
investigate whether an accurate value for the anomalous dimension of
the condensate can be extracted from topological observables. The rest of the paper
is organised as follows. In Sect.~\ref{methodology} we review the
formulation of the theory of interest on the lattice and we define the
observables related to topological quantities that are studied in this work. Our
numerical results are reported in Sect.~\ref{results}, with
discussions of methods for extracting the anomalous dimension of the
condensate delegated to Sect.~\ref{sect:scaling}. Finally,
Sect.~\ref{conclusions} summarises the main findings of our study.

\section{Methodology}
\label{methodology}
In this work, we study the theory on an four-dimensional Euclidean
spacetime lattice. The lattice discretisation used here
is described in~\cite{DelDebbio2008zf}. We recall
the main points below, referring to~\cite{DelDebbio2008zf} for further details. 

On the lattice, the dynamics of the gauge variables is
described by group elements in the fundamental representation living
on the links of the lattice. If $U_{\mu}(i)$ is the field living on
the link originating from the lattice point $i$ in the direction
$\hat{\mu}$, the plaquette operator $U_{\mu \nu}(i)$ is defined as  
\begin{equation}
U_{\mu \nu}(i) =
  U_{\mu}(i) U_{\nu}(i+\hat{\mu})
  U_{\mu}^{\dag}(i+\hat{\nu})U_{\nu}^{\dag}(i) \ .
\end{equation}
In $\su{N}$, the Wilson action for the gauge degrees of freedom of the theory
is then 
\begin{equation}
  S_{\mathrm{g}} = \beta \sum _{i, \mu, \nu} \left( 1 - \frac{1}{2N} \mbox{Tr}
    \left( U_{\mu \nu}(i) + U_{\mu \nu}^{\dag}(i) \right) \right) \ ,
\end{equation}
with $\beta = 1/g^2$ and $g$ the gauge coupling.
In the Wilson formulation, which is the discretisation used in our
work, the dynamics of the fermions is described
by the Dirac operator $M_{\alpha \beta}(ij)$, defined as
\begin{eqnarray}
\nonumber
 M_{\alpha \beta}(ij) &=& (m+4r)  \delta_{ij}
  \delta_{\alpha \beta} - \frac{1}{2} \left[\left(r - \gamma_{\mu}\right)_{\alpha
      \beta}U^{\mathrm{R}}_{\mu}(i) \delta_{i,j-\hat{\mu}} \right.\\
    &&\quad+ \left. \left(r  +
      \gamma_{\mu}\right)_{\alpha \beta} \left(U^{\mathrm{R}}_{\mu}(i)\right)^{\dag}\delta_{i,j+\hat{\mu}}
  \right] \ ,
\end{eqnarray}
where $i$ and $j$ are lattice sites and $\alpha$ and $\beta$ Dirac
indices. The superscript $\mathrm{R}$ indicates that the link variable have to
be taken in the representation of the fermions and $m$ is the bare
fermion mass. Since this fermion discretisation breaks chiral symmetry, the
chiral point has to be determined non-perturbatively in the numerical
simulation. A possible 
definition of this point uses the fermion mass $m_{\mathrm{PCAC}}$ defined
through the partially conserved axial current: the chiral limit is
attained at the point for which $m_{\mathrm{PCAC}}$ is zero.

With those definitions, the lattice discretised version of the $\su{2}$
gauge theory with two adjoint fermions we are studying is described by
the path integral
\begin{eqnarray}
 Z = \int \left( {\cal D} U_{\mu}(i)\right) (\det M(U_{\mu}))^{N_{\mathrm{f}}}
  e^{-S_{\mathrm{g}}} \ ,
\end{eqnarray}
where the link variables are $\su{2}$ matrices, $M$ is in the adjoint
representation and the number of flavours $N_{\mathrm{f}}$ is 2. 
A set of configurations that dominate the path integral can be
generated with Monte Carlo methods, and vacuum expectation values of
observables are computed as ensemble averages of the corresponding
operators over those configurations. These observables have a
relative statistical fluctuation that scales as $1/\sqrt{N_{\mathrm{s}}}$, where
$N_{\mathrm{s}}$ is the number of the generated configurations. Thus the numerical
error can be kept systematically under control by adjusting the size
of the sample.

The observables we have studied are described
in~\cite{smith-teper}, to which we refer for further details about the
observables and the model assumptions made to derive the relationships we
use in our work. Here we only summarise the main points. In the
continuum, the topological charge density can be expressed as 
\begin{equation}
Q(x) = \frac{1}{32\pi^2} \epsilon_{\mu\nu\rho\sigma} \tr
\left\{F_{\mu\nu}(x)F_{\rho\sigma}(x)\right\} \ .
\end{equation}
The total topological charge of a configuration can then be obtained
as the integral of this quantity over the spacetime.

The equivalent lattice topological charge density is then given by: 
\begin{equation}
\label{eq:td}
Q_\LL(i)= \frac{1}{32\pi^2} \epsilon_{\mu\nu\rho\sigma} \tr
\left\{U_{\mu\nu}(i)U_{\rho\sigma}(i)\right\} \ .
\end{equation}

For a smooth gauge field, this would have fluctuations of order $a^{2}$
as the lattice spacing $a$ is sent to zero;
however, realistic fields have ultraviolet fluctuations which will
completely dominate over the physics of interest in the continuum
limit. To mitigate this, a cooling process is
introduced~\cite{Teper1985rb}. The cooling process operates by
minimising the local action for each lattice site in turn. Successive
cooling sweeps will ``smooth out'' the fluctuations such that the
physics may be observed. A side effect however is that instantons may
be shrunk or may annihilate in instanton--anti-instanton pairs, thus
excessive cooling is to be avoided. The number of cooling sweeps
performed in a calculation results from the compromise between the
need to smooth out the configurations and the necessity of not losing
physical instantons by annihilation. The tuning of this parameter is
not critical, since there is a wide plateau where the physics can be
observed before cooling artefacts set in.

Once the topological charge density is known, various observables may be calculated. The topological susceptibility is defined as
\begin{equation}
\chi_\mathrm{T} = \frac{\left\langle Q_{\mathrm{T}}^2 \right\rangle - {\left\langle
      Q_{\mathrm{T}} \right\rangle}^2}{V} \ ,
\end{equation}
where $V$ is the lattice volume and $Q_{\mathrm{T}}$ is the total topological
charge, defined as
\begin{equation}
Q_{\mathrm{T}} = \sum_i Q_\LL(i) \ ,
\end{equation}
with $i$ running over all lattice points. Since in the continuum the
total topological charge is an integer, on the lattice $Q_{\mathrm{T}}$ is often
rounded to the nearest integer. Alternative discretisations can be
used. Lattice studies (e.g.~\cite{lucini-teper,Alles:1997nu}) show
that the particular definition of $Q_{\mathrm{T}}$ does not affect the results if
the calculation is done close to the continuum limit, as one would
have expected. 

The size of a given instanton may be calculated from the local maxima
of the absolute value of the topological charge density given in Eq.~(\ref{eq:td}) from the relation
\begin{equation}
q_\mathrm{peak} = \frac{6}{\pi^2\rho^4} \ ,
\end{equation}
which has been used to determine the instanton size distribution and the average instanton size.

\begin{figure}
  \includegraphics[width=0.95\columnwidth]{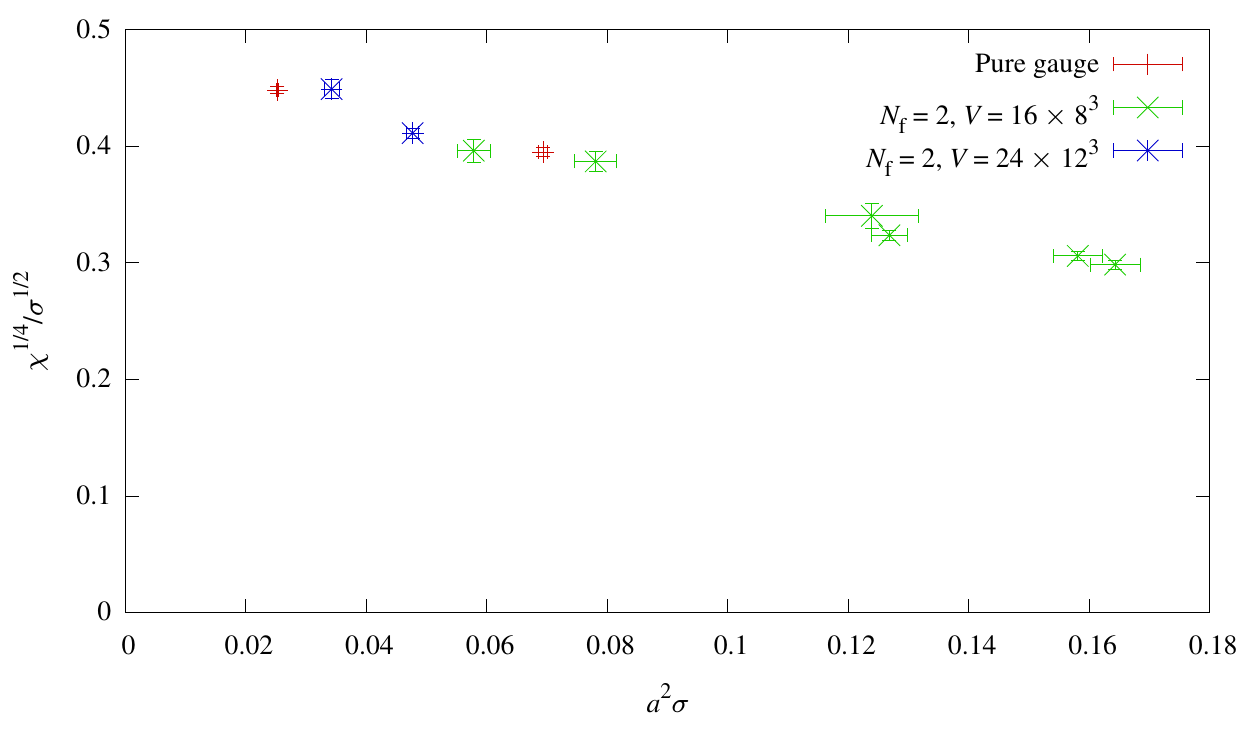}
  \caption{Variation of the quantity
  $\chi^\frac{1}{4}/\sqrt{\sigma}$ with $a^2 \sigma$. Smaller string
  tensions corresponds to smaller fermion masses.}
\label{fig:suscept}
\end{figure}

\section{Results}
\label{results}
Preexisting two-flavour configurations discussed
in~\cite{DelDebbio2010hx} (to which we refer for measurements of
spectral quantities and for a careful discussion of finite volume
artefacts) were used, at $\beta = 2.25$, with $V = 16\times 8^3$, $m =
-0.90$, $-0.75$, $-0.50$, $-0.25$, $0.00$, $0.25$, $0.50$ and $V = 
24\times 12^3$, $m = -1.00$, $-0.95$. 20 cooling sweeps were used for all
configurations.  For comparison with the Yang-Mills case, pure gauge
configurations were generated at $\beta = 2.4, V = 12^4$ and $\beta =
2.55, V = 20^4$; each update consisted of 1 heat bath and 4
over-relaxation steps, and measurements were taken every 10th
and 50th configuration on the smaller and larger lattice respectively.
At each $\beta$, the size of the lattice has been chosen
so that finite size effects for spectral observables are negligible,
with the linear size of the lattice (in physical units) being around
1.5 fm in both cases~\cite{lucini-teper} (for the conversion to fm, we
have assumed $\sqrt{\sigma} = 420$ MeV). Our numerical results have
been obtained with a bootstrap procedure over the data divided in bins
of 20 measurements each.

Before discussing our findings, it is important to note that, although
apparently similar from an operational point of view and for scope and
results, the comparison with the pure gauge results (which is an
important part of this work) is conceptually different from a
comparison with quenched data in QCD with heavy quarks: in the 
latter theory, a large quark mass only slightly modifies the string
tension, and its effect can be reabsorbed with a shift in
$\beta$. Here (see
Refs.~\cite{DelDebbio2009fd,DelDebbio2010hu,DelDebbio2010hx})
when the (small) fermion mass 
is varied towards the chiral limit, a variation of the value of the
string tension by an order of magnitude is induced on $\sigma$. Since
the lattice spacing is not changing ($\beta$ is kept fixed across the
various values of the mass), the interpretation (consistent with
calculations around the Caswell-Banks-Zaks~\cite{Caswell1974,Banks:1981nn} fixed point) is that it is the
scale of the long-distance effective Yang-Mills theory that changes as
a result of varying the mass. As a consequence, a comparison with the
Yang-Mills theory entails the matching of the quantity $a
\sqrt{\sigma}$. For this reason, even if in the
dynamical theory we are at fixed $\beta$, for the comparison one has to choose a
different value of the coupling in the pure gauge theory for each
value of the fermion mass in the dynamical theory. 

\begin{savenotes}
\begin{table}
\caption{Numerical results for $m_{PCAC}$, $\chi$ and
  $\overline{\rho}$ as a function of $m$ on lattices of size $2L \times L^3$. Note that we have removed the point at $m = 0.50$ from plots making use
of $\sigma$, owing to the unavailability of a reliable value of $\sigma$ for this data set.}
\label{tab:results}
\begin{tabular}{lllll}
\hline\noalign{\smallskip}
$a L$ & $m$ & $am_{\mathrm{PCAC}}$ & $a^4\chi$ & $\overline{\rho} / a$  \\
\noalign{\smallskip}\hline\noalign{\smallskip}
8 &	$-0.90$ &	0.4330(18) &	$8.23(17)\times {10}^{-5}$ &	4.12(67) \\
8 &	$-0.75$ &	0.5607(18) &	$1.365(23)\times {10}^{-4}$ &	4.02(51) \\
8 &	$-0.50$ &	0.7224(13) &	$1.764(28)\times {10}^{-4}$ &	3.96(41) \\
8 &	$-0.25$ &	0.8552(11) &	$2.059(33)\times {10}^{-4}$ &	3.88(37) \\
8 &	$\phantom{-}0.00$ &	0.9706(11) &	$2.188(37)\times {10}^{-4}$ &	3.86(35) \\
8 &	$\phantom{-}0.25$ &	1.07205(97) &	$2.141(35)\times {10}^{-4}$ &	3.85(34) \\
8 &	$\phantom{-}0.50$ &	1.16353(73) &	$2.198(36)\times {10}^{-4}$ &	3.84(34) \\
\noalign{\smallskip}\hline\noalign{\smallskip}
12 &	$-1.00$ &	0.33623(82) &	$6.502(93)\times {10}^{-5}$ &	6.18(94) \\
12 &	$-0.95$ &	0.39017(68) &	$4.772(81)\times {10}^{-5}$ &	5.73(67) \\
\noalign{\smallskip}\hline
\end{tabular}
\end{table}
\end{savenotes}

\begin{figure}
  \includegraphics[width=0.95\columnwidth]{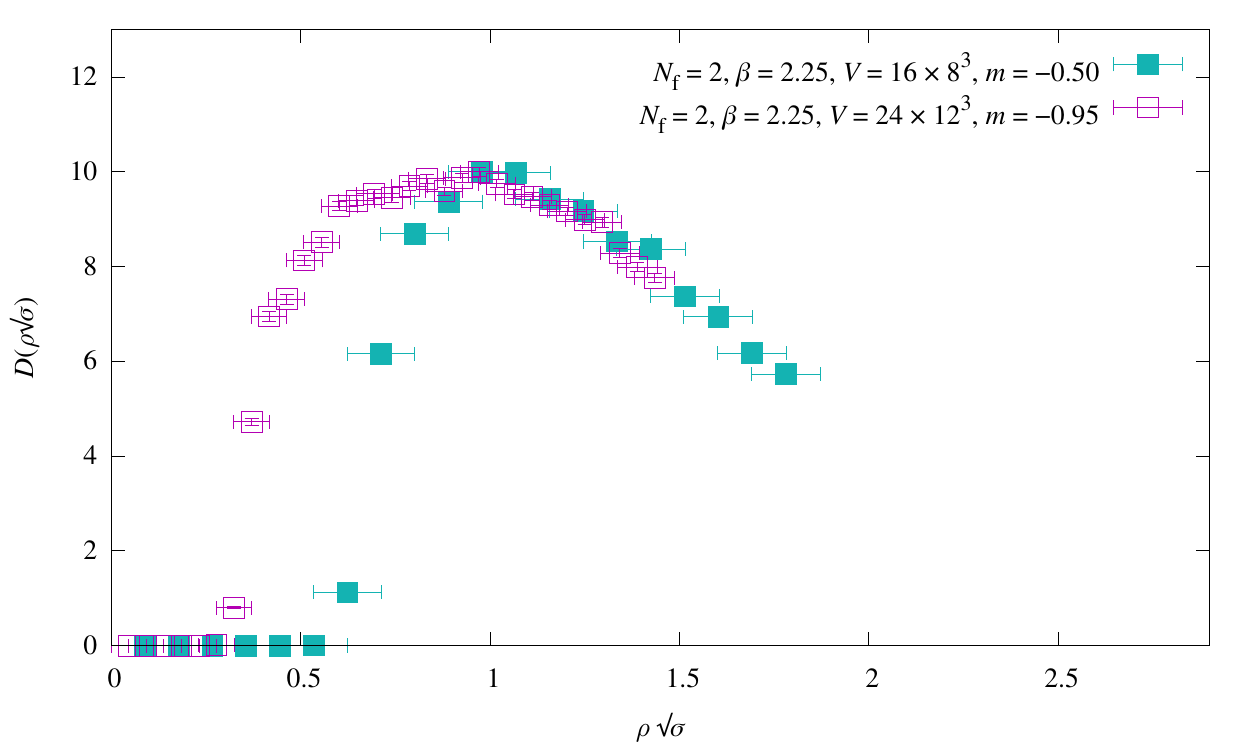}
  \caption{Comparison between the instanton size distributions for the
  dynamical case at the smaller and larger lattice volume, for the sets
  of bare parameters shown.}
\label{fig:rhodist-comparison}
\end{figure}

We start with the topological susceptibility. For this quantity, we
have used the integer definition of $Q_{\mathrm{T}}$, after having verified that
alternative definitions give compatible results. Our findings
(reported in table~\ref{tab:results} and plotted in figure~\ref{fig:suscept}) show
consistency between the pure gauge and the 2-flavour theory, as one
would have expected. As a check, we have verified that our pure gauge  
results are consistent with those presented in~\cite{lucini-teper}.  
\begin{figure}
  \includegraphics[width=0.95\columnwidth]{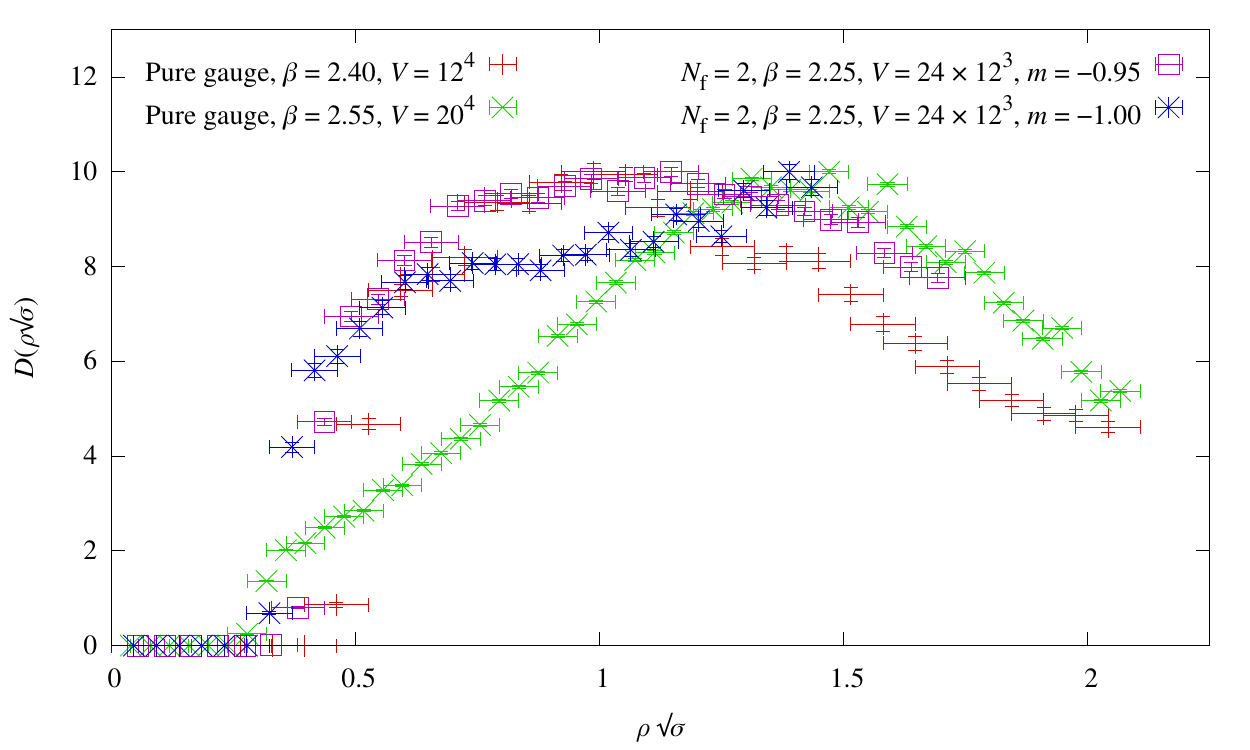}
  \caption{Instanton size distribution for the two-flavour configuration
  sets and for the pure gauge
  configuration sets at the values of the lattice parameters shown.}
\label{fig:rhodist-neat}
\end{figure}
\begin{figure}
  \includegraphics[width=0.95\columnwidth]{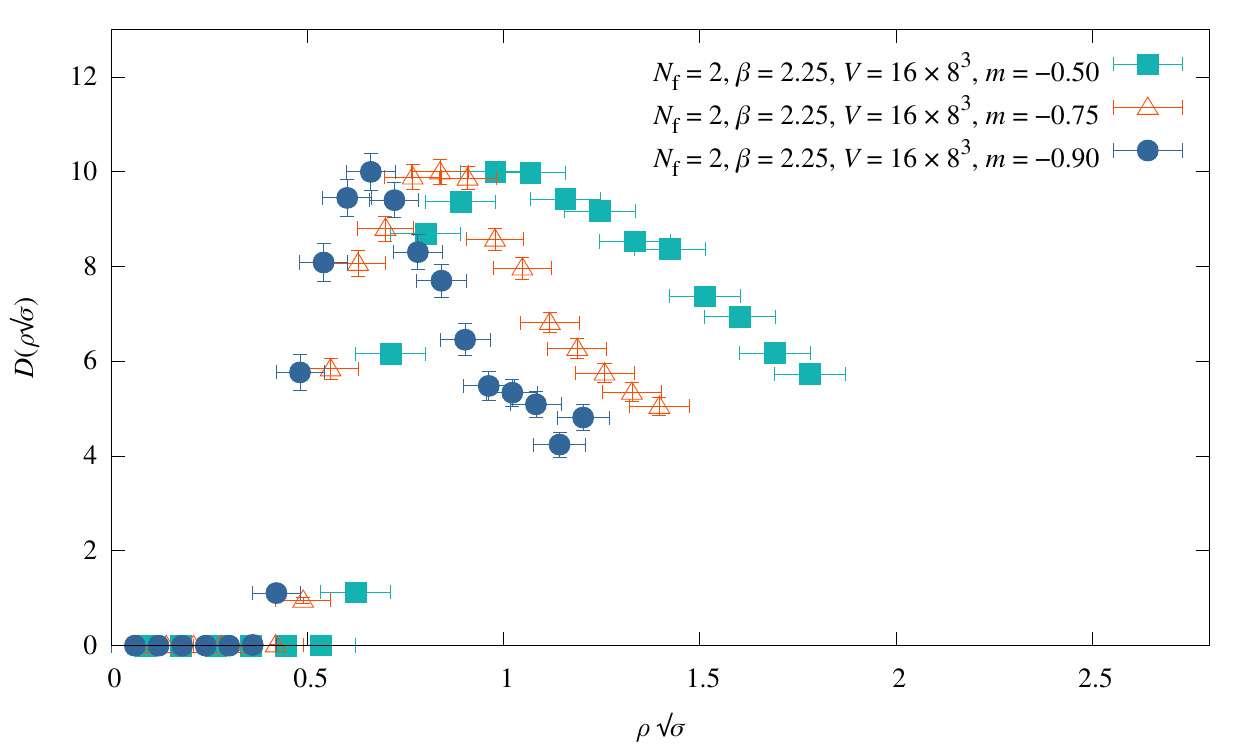}
  \caption{Instanton size distribution in the dynamical case on the
  smaller lattice at the parameters shown. The appearance of a peak at
  larger instanton size signals the onset of finite size artefacts.}
\label{fig:rhodist-finitesize}
\end{figure}
Although no sign of finite size effects can be seen in the topological
susceptibility, there are visible hints of them in the instanton size
distribution. 
We start our discussion by comparing this observable in the pure gauge case
and in the two flavour theory on the larger lattice\footnote{Note that
  even in the pure gauge case, contrary e.g. to glueball masses and to
  the topological susceptibility, the instanton size distribution at
  the needed values of $\beta$ has visible finite size effects. For this reason,
 statements involving this distribution should be regarded as
 qualitative rather than precise, quantitative characterisations of
 the system.}.  The instanton
size distribution (figure~\ref{fig:rhodist-neat}, showing the distributions of instantons of
physical size $\rho \sqrt{\sigma}$) has a similar behaviour,
characterised by a peak and a large size tail, whose end is possibly
affected by finite size artefacts. When going from larger to smaller
string tensions, the evolution of the shape of the distribution in the
dynamical case seems to be compatible with the quenched one. This behaviour is consistent with the
distribution found in studies of the same quantity in other
contexts~\cite{instanton-size-1,instanton-size-2,instanton-size-3}\footnote{We 
  remind the reader that the instanton size distribution, as measured in lattice simulations, is
  affected by cooling artefacts in the small size regime and by finite
  instanton size artefacts when the size becomes of the order of the
  lattice size $L$. Any physical statement concerning the instanton size
  distribution must concern the regime $1 \ll \rho \ll L$. For our
  range of parameters, this singles out a region around the peak of
  the distribution extending to a wide portion of the tail at larger
  $\rho$. The exact extent of this region would require a finite size
  and continuum study that are outside the scope of this work.}.  
However, on the smaller lattice, as the fermion mass is decreased,
value, the $N_{\mathrm{f}} = 2$ theory shows a sharper peak, a
position of the peak moved towards smaller sizes and a
longer tail (figure~\ref{fig:rhodist-finitesize}). This signals the
onset of finite volume artefacts. If one avoids the region in which
lattice artefacts are dominant, when rescaled in units of
$\sqrt{\sigma}$, the instanton size distribution on different lattice
volumes shows the same large-size behaviour, while at smaller
instanton sizes the distribution on the smaller lattice is suppressed
(figure ~\ref{fig:rhodist-comparison}). We ascribe this suppression to
more prominent cooling artefacts that cause instantons to shrink below
the lattice spacing on smaller lattices.

\section{Anomalous dimension and scaling}
\label{sect:scaling}
In a conformal gauge theory, for which the fermion mass is the only
relevant coupling at the infrared fixed point, at large distances and in the infinite volume limit all
observables scale with an exponent related to their na\"ive mass
dimension and to the anomalous dimension of the condensate. Let us
consider a multiplicatively renormalised mass $m$ such that the chiral
limit is attained at $m = 0$ (like for instance the PCAC mass).
If $\gamma^{\ast}$ is the anomalous dimension of the condensate, at
leading order in $m$ an observable $O$ with mass dimension $d_O$
scales as 
\begin{equation}
\label{eq:scal:1}
O \propto m^{d_O/(1 + \gamma^{\ast})} \ .
\end{equation}
The argument parallels the derivation of scaling relations of
statistical systems at criticality and assumes that the correlation
length of the system is the only relavant
scale~\cite{DeGrand2009mt}. This assumption is known
as the hyperscaling hypothesis.

Infinite volume scaling relations emerge as the large size limit
of finite size scaling, which in turn can be derived by adding the size
of the system, $L$, to the set of relevant
couplings~\cite{Lucini2009an,DelDebbio2010hx,DelDebbio2010jy,DelDebbio2010ze}. Hence,
for a finite system of size $L$, Eq.~(\ref{eq:scal:1}) becomes
\begin{equation}
\label{eq:scal:2}
O L^{d_O} = F_O (L m^{1/(1 + \gamma^{\ast})}) \ ,
\end{equation}
where $F_O$ is a universal function of of the scaling variable $x = L
m^{1/(1 + \gamma^{\ast})}$. At large $L$ and small $x$, 
\begin{equation}
\label{eq:scal:3} 
F_O = x^{d_O} + \dots \ ,
\end{equation}
and Eq.~(\ref{eq:scal:2}) reduced to Eq.~(\ref{eq:scal:1}). 
\begin{figure}
  \includegraphics[width=0.95\columnwidth]{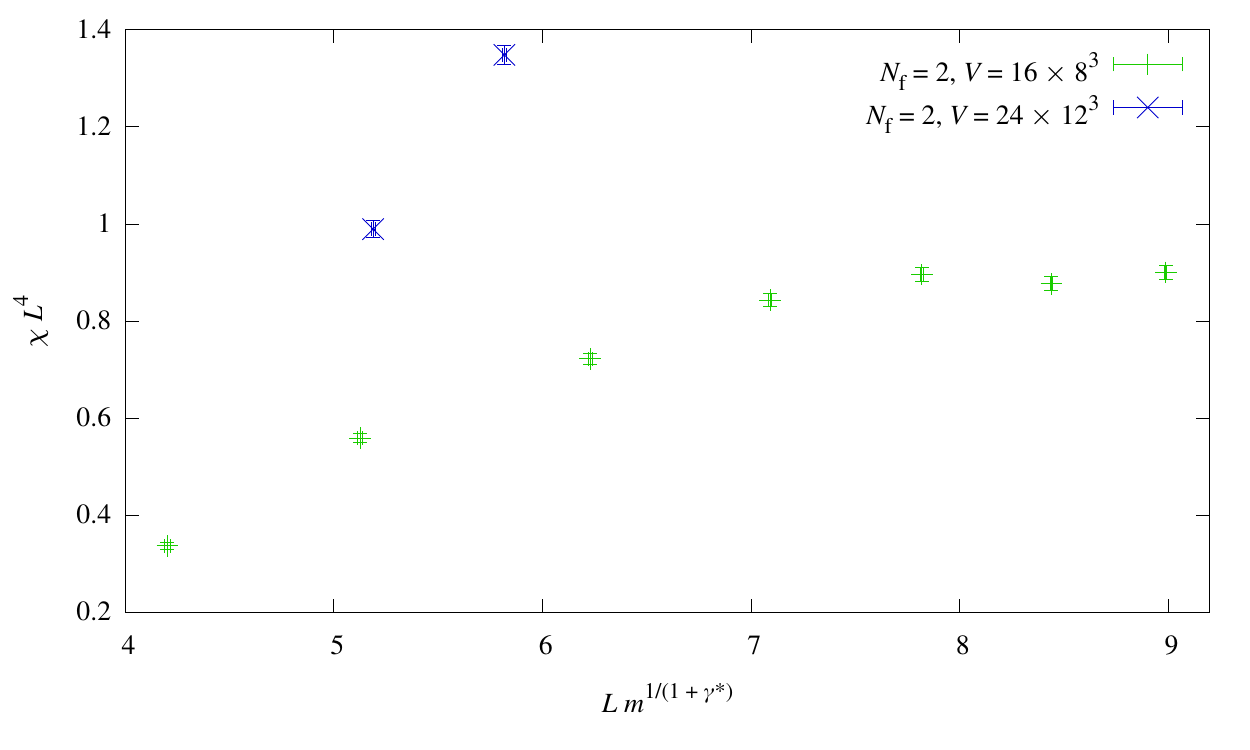}
  \caption{Rescaling of the topological susceptibility data according
    to Eq.~(\ref{eq:1.2}) with $\gamma^{\ast} = 0.3$.}
\label{fig:susc:scaling} 
\end{figure}
\begin{figure}
  \includegraphics[width=0.95\columnwidth]{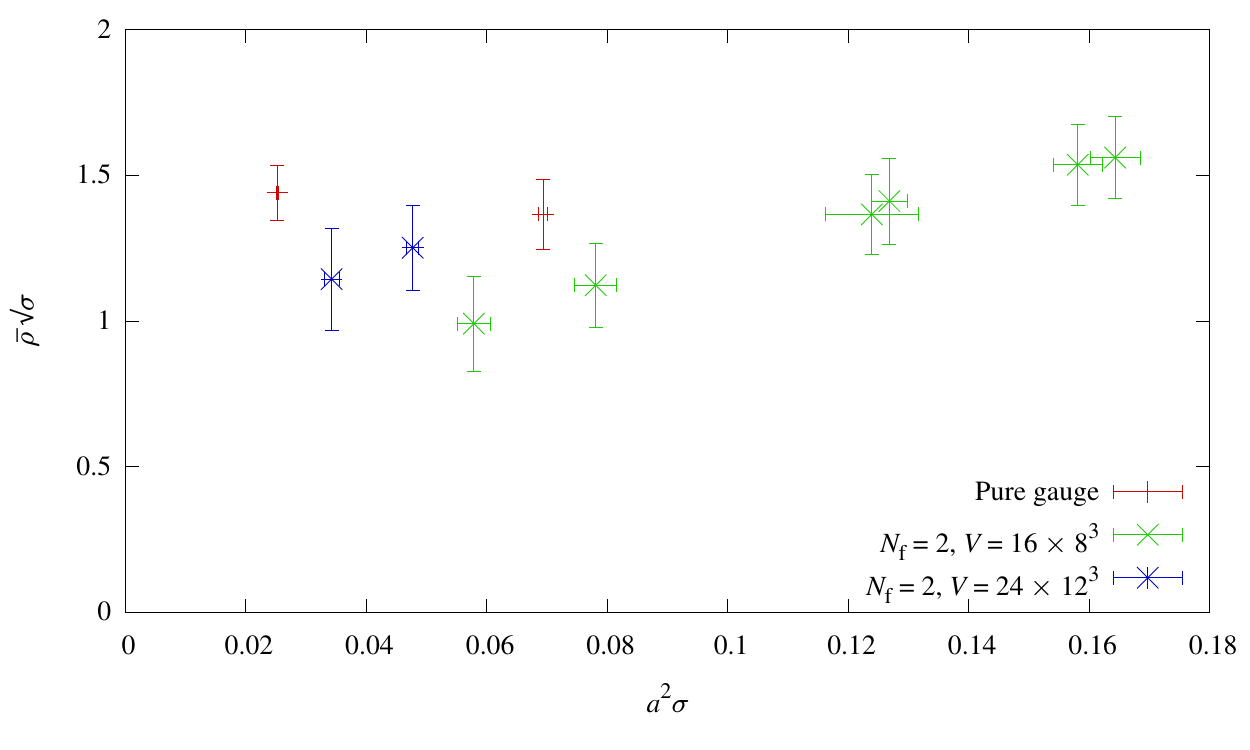}
  \caption{Variation of the average instanton size with $\sigma$.}
\label{fig:rho} 
\end{figure}
As a consequence of scaling and finite volume scaling arguments,
under the assumption that the marginal $\theta$-term operator does not 
generate sizeable scaling violations, in the large volume and small mass
regime, the topological susceptibility $\chi$ will show the scaling
behaviour 
\begin{equation}
\label{eq:1.1}
\chi \propto m^{\frac{4}{1 + \gamma^{\ast}}} \ .
\end{equation}
At finite but large volume $L^4$, this
relationship becomes
\begin{equation}
\label{eq:1.2}
\chi = L^{-4} F_{\chi} (L m^{1/(1 + \gamma^{\ast})}) \ ,
\end{equation}
with $F_\chi$ universal function of $L m^{1/(1 + \gamma^{\ast})}$ in
the limits $L \to \infty$ and $m\to 0$. 
Analogously, the instanton size distribution follows the scaling and
finite size scaling laws
\begin{eqnarray}
\label{eq:scal:9}
\overline{\rho} &=& A m^{- 1/(1 + \gamma^{\ast})} \ , \\ 
\overline{\rho} &=& L F_{\overline{\rho}} (L m^{1/(1 +
  \gamma^{\ast})}) \ ,  
\end{eqnarray}
respectively, with $A$ in Eq.~(\ref{eq:scal:9}) a constant.

We performed a fit of the topological susceptibility data according to
Eqs.~(\ref{eq:1.1})~and~(\ref{eq:1.2}), but we did not see the
expected scaling. Even putting a reasonable bound on $\gamma^{\ast}$ proves
to be hard with the current data: adjusting by hand the value of
$\gamma^{\ast}$ in the range 0.0--1.0 did not result in any universal
behaviour as a function of the scaling variable. An example is
provided in figure~\ref{fig:susc:scaling}. 

The average instanton size is reported in
table~\ref{tab:results} and plotted in figure~\ref{fig:rho}. This
quantity shows qualitative scaling and a behaviour compatible with the
pure gauge case. However, the relatively large error on $\overline{\rho}$ does not
allow to extract an accurate value for the anomalous dimension: an
acceptable scaling is obtained for $0 \le \gamma^{\ast} \le 2$. 

While the lack of precision in the determination of $\gamma^{\ast}$
from $\overline{\rho}$ is due to the intrinsical difficulties in measuring
reliably this observable, the most plausible explanation for the
apparent absence of scaling in $\chi$ is that larger
lattices are required in order for the scaling behaviour to
manifest. Although this could have been expected from previous
studies, in~\cite{Patella2012da} it was shown an 
example of an observable that has precocious scaling. Our analysis
shows that this is not the case for the topological susceptibility.
Generating configurations in the scaling region of gluonic
observables requires a big computational effort, which is outside the
scope of this work.  Currently, a calculation in this direction is
being performed~\cite{DelDebbio2011kp}. These newly generated configurations will
allow us to investigate the scaling of the topological susceptibility
as the mass goes to zero in the appropriate range of volumes and then
to answer the question of whether the topological suceptibility is an
efficient observable to extract the anomalous dimension.

\section{Conclusions}
\label{conclusions}
In this work, we performed the first numerical investigation of the behaviour of
instanton-related quantities in a mass-deformed infrared conformal
gauge theory. For the theory we have studied, $\su{2}$ gauge theory with
two dynamical adjoint Dirac fermions, we found that the behaviour of
the topological susceptibility supports the scenario in which
the infrared dynamics is dominated by gluonic quantities, with
the fermions being more
massive~\cite{Miransky1998dh,DelDebbio2009fd} (see
also~\cite{Sannino:2008pz}). Moreover, the instanton size distribution
provides a good indication of finite volume
effects. We plan to extend this study to larger lattices, in order to
check whether the topological susceptibility obeys the expected scaling
with the mass and whether we can extract from it the anomalous
dimension of the condensate. Finally, it would be interesting to apply
those techniques to other theories that could provide a realisation of
the walking scenario, like e.g. $\su{3}$ gauge theory with two sextet
fermions~\cite{DeGrand2012yq,Holland2012fu}.

%
%

\begin{acknowledgements}
We thank A. Patella and A. Rago for discussions.
This work was done as part of the UKQCD collaboration and the DiRAC
Facility jointly funded by STFC, the Large Facilities Capital Fund of
BIS and Swansea University. We are indebted to L. Del Debbio,
A. Patella, C. Pica and A. Rago, who made available to us the
configurations discussed in~\cite{DelDebbio2010hu,DelDebbio2010hx}. EB
is supported by STFC. BL is supported by the Royal Society and by
STFC.
\end{acknowledgements}

\bibliographystyle{spphys}       
\bibliography{topologymwt.bib}   

\begin{thebibliography}{10}
\providecommand{\url}[1]{{#1}}
\providecommand{\urlprefix}{URL }
\expandafter\ifx\csname urlstyle\endcsname\relax
  \providecommand{\doi}[1]{DOI \discretionary{}{}{}#1}\else
  \providecommand{\doi}{DOI \discretionary{}{}{}\begingroup
  \urlstyle{rm}\Url}\fi

\bibitem{atlas2012gk}
G.~Aad, et~al., Phys.Lett.B  (2012).
\newblock \doi{10.1016/j.physletb.2012.08.020}

\bibitem{cms2012gu}
S.~Chatrchyan, et~al., Phys.Lett.B  (2012).
\newblock \doi{10.1016/j.physletb.2012.08.021}

\bibitem{Weinberg1975gm}
S.~Weinberg, Phys. Rev. \textbf{D13}, 974 (1976).
\newblock \doi{10.1103/PhysRevD.13.974}

\bibitem{Susskind1978ms}
L.~Susskind, Phys. Rev. \textbf{D20}, 2619 (1979).
\newblock \doi{10.1103/PhysRevD.20.2619}

\bibitem{Eichten1979ah}
E.~Eichten, K.D. Lane, Phys. Lett. \textbf{B90}, 125 (1980).
\newblock \doi{10.1016/0370-2693(80)90065-9}

\bibitem{Holdom1981rm}
B.~Holdom, Phys.Rev. \textbf{D24}, 1441 (1981).
\newblock \doi{10.1103/PhysRevD.24.1441}

\bibitem{Holdom1984sk}
B.~Holdom, Phys. Lett. \textbf{B150}, 301 (1985).
\newblock \doi{10.1016/0370-2693(85)91015-9}

\bibitem{Yamawaki1985zg}
K.~Yamawaki, M.~Bando, K.i. Matumoto, Phys. Rev. Lett. \textbf{56}, 1335
  (1986).
\newblock \doi{10.1103/PhysRevLett.56.1335}

\bibitem{Appelquist1986an}
T.W. Appelquist, D.~Karabali, L.C.R. Wijewardhana, Phys. Rev. Lett.
  \textbf{57}, 957 (1986).
\newblock \doi{10.1103/PhysRevLett.57.957}

\bibitem{Miransky1996pd}
V.~Miransky, K.~Yamawaki, Phys.Rev. \textbf{D55}, 5051 (1997).
\newblock \doi{10.1103/PhysRevD.56.3768, 10.1103/PhysRevD.55.5051}

\bibitem{Dietrich2006cm}
D.D. Dietrich, F.~Sannino, Phys.Rev. \textbf{D75}, 085018 (2007).
\newblock \doi{10.1103/PhysRevD.75.085018}

\bibitem{Dietrich2006ck}
D.D. Dietrich, F.~Sannino, Phys. Rev. \textbf{D75}, 085018 (2007).
\newblock \doi{10.1103/PhysRevD.75.085018}

\bibitem{Hill2002ap}
C.T. Hill, E.H. Simmons, Phys. Rept. \textbf{381}, 235 (2003).
\newblock \doi{10.1016/S0370-1573(03)00140-6}

\bibitem{Piai2010ma}
M.~Piai, Adv. High Energy Phys. \textbf{2010} (4302).
\newblock \doi{10.1155/2010/464302}

\bibitem{Andersen2011yj}
J.~Andersen, O.~Antipin, G.~Azuelos, L.~Del~Debbio, E.~Del~Nobile, et~al.,
  Eur.Phys.J.Plus \textbf{126}, 81 (2011).
\newblock \doi{10.1140/epjp/i2011-11081-1}

\bibitem{Nunez2008wi}
C.~Nunez, I.~Papadimitriou, M.~Piai, Int. J. Mod. Phys. \textbf{A25}, 2837
  (2010).
\newblock \doi{10.1142/S0217751X10049189}

\bibitem{Elander2009pk}
D.~Elander, C.~Nunez, M.~Piai, Phys. Lett. \textbf{B686}, 64 (2010).
\newblock \doi{10.1016/j.physletb.2010.02.023}

\bibitem{Kutasov2012uq}
D.~Kutasov, J.~Lin, A.~Parnachev, Nucl.Phys. \textbf{B863}, 361 (2012).
\newblock \doi{10.1016/j.nuclphysb.2012.05.025}

\bibitem{Levkov2012yk}
D.~Levkov, V.~Rubakov, S.~Troitsky, Y.~Zenkevich, arXiv:1201.6368  (2012)

\bibitem{Anguelova2012ka}
L.~Anguelova, P.~Suranyi, L.R. Wijewardhana, Nucl.Phys. \textbf{B862}, 671
  (2012).
\newblock \doi{10.1016/j.nuclphysb.2012.05.005}

\bibitem{Clark2012pm}
T.~Clark, S.~Love, T.~ter Veldhuis, arXiv:1208.0817  (2012)

\bibitem{Neil2012cb}
E.T. Neil, PoS \textbf{LATTICE2011}, 009 (2011)

\bibitem{Sannino:2004qp}
F.~Sannino, K.~Tuominen, Phys.Rev. \textbf{D71}, 051901 (2005).
\newblock \doi{10.1103/PhysRevD.71.051901}

\bibitem{Dietrich:2005jn}
D.D. Dietrich, F.~Sannino, K.~Tuominen, Phys.Rev. \textbf{D72}, 055001 (2005).
\newblock \doi{10.1103/PhysRevD.72.055001}

\bibitem{Foadi:2007ue}
R.~Foadi, M.T. Frandsen, T.A. Ryttov, F.~Sannino, Phys.Rev. \textbf{D76},
  055005 (2007).
\newblock \doi{10.1103/PhysRevD.76.055005}

\bibitem{atlas-mwt}
G.~Aad, et~al., arXiv:1209.2535  (2012)

\bibitem{Luty2004ye}
M.A. Luty, T.~Okui, JHEP \textbf{0609}, 070 (2006).
\newblock \doi{10.1088/1126-6708/2006/09/070}

\bibitem{Catterall2007yx}
S.~Catterall, F.~Sannino, Phys. Rev. \textbf{D76}, 034504 (2007).
\newblock \doi{10.1103/PhysRevD.76.034504}

\bibitem{DelDebbio2008zf}
L.~Del~Debbio, A.~Patella, C.~Pica, Phys. Rev. \textbf{D81}, 094503 (2010).
\newblock \doi{10.1103/PhysRevD.81.094503}

\bibitem{Catterall2008qk}
S.~Catterall, J.~Giedt, F.~Sannino, J.~Schneible, JHEP \textbf{11}, 009 (2008).
\newblock \doi{10.1088/1126-6708/2008/11/009}

\bibitem{Hietanen2008mr}
A.J. Hietanen, J.~Rantaharju, K.~Rummukainen, K.~Tuominen, JHEP \textbf{0905},
  025 (2009).
\newblock \doi{10.1088/1126-6708/2009/05/025}

\bibitem{Hietanen2009az}
A.J. Hietanen, K.~Rummukainen, K.~Tuominen, Phys. Rev. \textbf{D80}, 094504
  (2009).
\newblock \doi{10.1103/PhysRevD.80.094504}

\bibitem{Bursa2009we}
F.~Bursa, L.~Del~Debbio, L.~Keegan, C.~Pica, T.~Pickup, Phys. Rev.
  \textbf{D81}, 014505 (2010).
\newblock \doi{10.1103/PhysRevD.81.014505}

\bibitem{DeGrand2011qd}
T.~DeGrand, Y.~Shamir, B.~Svetitsky, Phys.Rev. \textbf{D83}, 074507 (2011).
\newblock \doi{10.1103/PhysRevD.83.074507}

\bibitem{Catterall2011zf}
S.~Catterall, L.~Del~Debbio, J.~Giedt, L.~Keegan, Phys.Rev. \textbf{D85},
  094501 (2012).
\newblock \doi{10.1103/PhysRevD.85.094501}

\bibitem{Lucini2009an}
B.~Lucini, Phil.Trans.Roy.Soc.Lond. \textbf{A368}, 3657 (2010).
\newblock \doi{10.1098/rsta.2010.0030}

\bibitem{DelDebbio2009fd}
L.~Del~Debbio, B.~Lucini, A.~Patella, C.~Pica, A.~Rago, Phys.Rev. \textbf{D80},
  074507 (2009).
\newblock \doi{10.1103/PhysRevD.80.074507}

\bibitem{DelDebbio2010hu}
L.~Del~Debbio, B.~Lucini, A.~Patella, C.~Pica, A.~Rago, Phys.Rev. \textbf{D82},
  014509 (2010).
\newblock \doi{10.1103/PhysRevD.82.014509}

\bibitem{DelDebbio2010hx}
L.~Del~Debbio, B.~Lucini, A.~Patella, C.~Pica, A.~Rago, Phys. Rev. D
  \textbf{82}, 014510 (2010).
\newblock \doi{10.1103/PhysRevD.82.014510}.
\newblock \urlprefix\url{http://link.aps.org/doi/10.1103/PhysRevD.82.014510}

\bibitem{Giedt2012rj}
J.~Giedt, E.~Weinberg, Phys.Rev. \textbf{D85}, 097503 (2012).
\newblock \doi{10.1103/PhysRevD.85.097503}

\bibitem{Patella2012da}
A.~Patella, Phys.Rev. \textbf{D86}, 025006 (2012).
\newblock \doi{10.1103/PhysRevD.86.025006}

\bibitem{Karavirta2011mv}
T.~Karavirta, A.~Mykkanen, J.~Rantaharju, K.~Rummukainen, K.~Tuominen, JHEP
  \textbf{1106}, 061 (2011).
\newblock \doi{10.1007/JHEP06(2011)061}

\bibitem{Bursa2011ru}
F.~Bursa, L.~Del~Debbio, D.~Henty, E.~Kerrane, B.~Lucini, et~al., Phys.Rev.
  \textbf{D84}, 034506 (2011).
\newblock \doi{10.1103/PhysRevD.84.034506}

\bibitem{DelDebbio2011kp}
L.~Del~Debbio, B.~Lucini, A.~Patella, C.~Pica, A.~Rago, PoS
  \textbf{LATTICE2011}, 084 (2011)

\bibitem{Nogradi2012dj}
D.~Nogradi, JHEP \textbf{1205}, 089 (2012).
\newblock \doi{10.1007/JHEP05(2012)089}

\bibitem{deForcrand2012se}
P.~de~Forcrand, M.~Pepe, U.J. Wiese, arXiv:1204.4913  (2012)

\bibitem{Delia2012vv}
M.~D'Elia, F.~Negro, Phys.Rev.Lett. \textbf{109}, 072001 (2012).
\newblock \doi{10.1103/PhysRevLett.109.072001}

\bibitem{Miransky1998dh}
V.~Miransky, Phys.Rev. \textbf{D59}, 105003 (1999).
\newblock \doi{10.1103/PhysRevD.59.105003}

\bibitem{Sannino:2008pz}
F.~Sannino, Phys.Rev. \textbf{D80}, 017901 (2009).
\newblock \doi{10.1103/PhysRevD.80.017901}

\bibitem{smith-teper}
D.A. Smith, M.J. Teper, Phys. Rev. D \textbf{58}, 014505 (1998).
\newblock \doi{10.1103/PhysRevD.58.014505}.
\newblock \urlprefix\url{http://link.aps.org/doi/10.1103/PhysRevD.58.014505}

\bibitem{Teper1985rb}
M.~Teper, Phys.Lett. \textbf{B162}, 357 (1985).
\newblock \doi{10.1016/0370-2693(85)90939-6}

\bibitem{lucini-teper}
B.~Lucini, M.~Teper, JHEP \textbf{0106}, 050 (2001).
\newblock \doi{10.1088/1126-6708/2001/06/050}

\bibitem{Alles:1997nu}
B.~Alles, M.~D'Elia, A.~Di~Giacomo, R.~Kirchner, Phys.Rev. \textbf{D58}, 114506
  (1998).
\newblock \doi{10.1103/PhysRevD.58.114506}

\bibitem{Caswell1974}
W.E. Caswell, Phys. Rev. Lett. \textbf{33}, 244 (1974).
\newblock \doi{10.1103/PhysRevLett.33.244}.
\newblock \urlprefix\url{http://link.aps.org/doi/10.1103/PhysRevLett.33.244}

\bibitem{Banks:1981nn}
T.~Banks, A.~Zaks, Nucl.Phys. \textbf{B196}, 189 (1982).
\newblock \doi{10.1016/0550-3213(82)90035-9}

\bibitem{instanton-size-1}
M.~Garc\'ia~P\'erez, T.G. Kov\'acs, P.~van Baal, Phys.Lett. \textbf{B472}, 295
  (2000).
\newblock \doi{10.1016/S0370-2693(99)01451-3}

\bibitem{instanton-size-2}
C.~Michael, P.~Spencer, Phys.Rev. \textbf{D52}, 4691 (1995).
\newblock \doi{10.1103/PhysRevD.52.4691}

\bibitem{instanton-size-3}
S.~Hands, P.~Kenny, Phys.Lett. \textbf{B701}, 373 (2011).
\newblock \doi{10.1016/j.physletb.2011.05.065}

\bibitem{DeGrand2009mt}
T.~DeGrand, A.~Hasenfratz, Phys. Rev. \textbf{D80}, 034506 (2009).
\newblock \doi{10.1103/PhysRevD.80.034506}

\bibitem{DelDebbio2010jy}
L.~Del~Debbio, R.~Zwicky, Phys.Lett. \textbf{B700}, 217 (2011).
\newblock \doi{10.1016/j.physletb.2011.04.059}

\bibitem{DelDebbio2010ze}
L.~Del~Debbio, R.~Zwicky, Phys. Rev. \textbf{D82}, 014502 (2010).
\newblock \doi{10.1103/PhysRevD.82.014502}

\bibitem{DeGrand2012yq}
T.~DeGrand, Y.~Shamir, B.~Svetitsky, arXiv:1201.0935  (2012)

\bibitem{Holland2012fu}
K.~Holland, Z.~Fodor, J.~Kuti, D.~Nogradi, C.~Schroeder, et~al.,
  arXiv:1209.0391  (2012)

\end{thebibliography}

\end{document}